\documentclass{PoS}
\usepackage{graphicx}
\usepackage[fleqn]{amsmath}
\usepackage{amssymb}
\newcommand{\be}{\begin{equation}}
\newcommand{\ee}{\end{equation}}
\newcommand{\bea}{\begin{eqnarray}}
\newcommand{\eea}{\end{eqnarray}}
\newcommand{\<}{\langle}
\renewcommand{\>}{\rangle}
\renewcommand{\[}{\langle\!\langle}
\renewcommand{\]}{\rangle\!\rangle}

\newcommand{\nn}{\nonumber}

\title{%
\begin{picture}(0,0)(0,0)%
\put(0,75){\makebox(0,0)[l]{\textnormal{\normalsize MKPH-T-07-15}}}%
\end{picture}%
Relativistic corrections to the static potential at
$O(1/m)$ and $O(1/m^2)$
}

\ShortTitle{
Relativistic corrections to the static potential at
$O(1/m)$ and $O(1/m^2)$
}

\author{Yoshiaki Koma\footnote{Speaker of 
``Determination of the velocity-dependent potentials
at $O(1/m^2)$''}%
\\
Numazu College of Technology,
Numazu 410-8501, Japan\\
E-mail: \email{koma@numazu-ct.ac.jp}}

\author{Miho Koma\footnote{Speaker of 
``Relativistic correction to the static potential at $O(1/m)$''}\\
Institut f\"ur Kernphysik, Johannes Gutenberg-Universit\"at Mainz,
D-55099 Mainz, Germany\\
E-mail: \email{mkoma@kph.uni-mainz.de}}

\author{Hartmut Wittig\\
Institut f\"ur Kernphysik, Johannes Gutenberg-Universit\"at Mainz,
D-55099 Mainz, Germany\\
E-mail: \email{wittig@kph.uni-mainz.de}}

\abstract{
We investigate the relativistic corrections to the
static potential, i.e. the $O(1/m)$ potential and the 
$O(1/m^{2})$ velocity-dependent potentials,
in SU(3) lattice gauge theory.
They are important  ingredients of 
potential nonrelativistic QCD for heavy quarkonium.
Utilizing the multi-level algorithm,
we obtain remarkably clean signals of 
these potentials up to $r=0.9$~fm.
We observe long range nonperturbative contributions
to these corrections.
}

\FullConference{The XXV International Symposium on
Lattice Field Theory\\
July 30 - August 4 2007\\
Regensburg, Germany}

\begin{document}

\section{Introduction}

A possible strategy of studying heavy quarkonium
in QCD is to employ potential nonrelativistic QCD
(pNRQCD)~\cite{Brambilla:1999xf,Brambilla:2000gk,%
Pineda:2000sz,Brambilla:2004jw},
which is obtained by integrating out the scale above
the heavy quark mass $m\gg \Lambda_{\rm QCD}$
and the scale $mv$, where $v$ is quark velocity.
The effective hamiltonian of pNRQCD 
up to $O(1/m^2)$~\cite{Pineda:2000sz} is then given by
\bea
H &=&
\frac{\vec{p}_{1}^{\;2}}{2m_{1}}
+\frac{\vec{p}_{2}^{\;2}}{2m_{2}}
+V^{(0)}(r) 
+\frac{1}{m_{1}}V^{(1,0)}(r)
+\frac{1}{m_{2}}V^{(0,1)}(r) \nn\\
&&
+\frac{1}{m_{1}^2}V^{(2,0)}(r)
+\frac{1}{m_{2}^2}V^{(0,2)}(r)
+\frac{1}{m_{1}m_{2}}V^{(1,1)}(r)
+O(1/m^3) \;,
\label{eqn:pnrqcd-hamiltonian}
\eea
where $m_{1}$ and $m_{2}$ denote the masses 
of quark and antiquark, 
placed at $\vec{r}_{1}$ and $\vec{r}_{2}$, respectively.
The static inter-quark potential 
$V^{(0)}(r \equiv |\vec{r}_{1}-\vec{r}_{2}|)$ 
emerges, accompanied by relativistic corrections classified 
in powers of~$1/m$.
The potentials $V^{(1,0)}(r)=V^{(0,1)}(r)~(\equiv V^{(1)}(r))$
are the corrections at $O(1/m)$.
The potentials $V^{(2,0)}(r)$, $V^{(0,2)}(r)$, and 
$V^{(1,1)}(r)$ are the corrections at $O(1/m^2)$, which
contain the leading order spin-dependent 
potentials~\cite{Eichten:1979pu,Eichten:1980mw,Gromes:1983pm}
and the velocity-dependent 
potentials~\cite{Barchielli:1986zs,Barchielli:1988zp}.
Spin-dependent potentials
are relevant to describing the fine and hyper-fine splitting 
of heavy quarkonium spectra.
Once these potentials are determined,
various properties of heavy quarkonium,
not only the ground state but also excited states,
e.g. full spectrum and wave functions,
can be investigated systematically 
by solving the Schr\"odinger equation.

\par
One may rely on perturbation theory to 
determine these potentials in the short-distance region.
For instance, perturbative studies of the $O(1/m)$ potential
yield $V^{(1)}(r)= - C_{F}C_{A} \alpha_{s}^{2}/(4 r^2)$%
~\cite{Melnikov:1998pr,Hoang:1998uv,Brambilla:2000gk},
where $C_{F}=4/3$ and $C_{A}=3$ are the Casimir charges of the 
fundamental and adjoint representations, respectively,
and $\alpha_{s}=g^2/(4\pi)$ the strong coupling 
(for the expression beyond leading-order perturbation theory, 
see~\cite{Brambilla:1999xj}).
However, since the binding energy of a quark and an antiquark
is typically of the scale~$mv^2$, which can be of the same order 
as $\Lambda_{\rm QCD}$ 
due to the nonrelativistic nature of the system, $v\ll 1$,
as well as the fact that perturbation theory
cannot incorporate quark confinement,
it is essential to determine the potentials
nonperturbatively.

\par
Monte Carlo simulations of lattice QCD offer a powerful
tool for the nonperturbative determination of the potentials.
The static potential 
can easily be determined from the expectation value of 
the $r\times t_{w}$ rectangular Wilson loop 
by $V^{(0)}(r)= -\lim_{t_{w} \to \infty}(1/t_{w})
\ln \< W(r,t_{w}) \>$.
The result is well parametrized by the Coulomb plus
linear confining term, 
\bea
V^{(0)}(r) = -\frac{c}{r} +\sigma r +\mu \;,
\label{eqn:pot-0}
\eea
where $c$ denotes the Coulombic
coefficient, $\sigma$ the string tension and $\mu$ a constant term. 
Recently, we investigated the $O(1/m)$ 
potential~\cite{Koma:2006si} and 
the $O(1/m^2)$ spin-dependent 
potentials~\cite{Koma:2005nq,Koma:2006fw}
on the lattice utilizing a new method, and 
obtained remarkably clean signals up to distances
of around 0.6~fm. We then observed a certain deviation from the
perturbative potentials at intermediate distances.

\par
In this presentation we further investigate the relativistic 
corrections to the static potential.
In particular, we aim to clarify the long-distance 
behavior of the $O(1/m)$ potential, and to 
determine the $O(1/m^2)$ velocity-dependent potentials.
One attempt to determine the velocity dependent potentials
on the lattice was published a decade ago~\cite{Bali:1997am},
but the signal was lost due to large statistical errors.

\section{Spectral representation of 
the $O(1/m)$ and $O(1/m^2)$ potentials}

\par
According to pNRQCD, the $O(1/m)$ and $O(1/m^2)$ potentials
can generally be expressed by
the {\em matrix elements} and the {\em energy gaps} 
of the spectral representation of 
the field strength correlators on the 
quark-antiquark source~\cite{Brambilla:2000gk,Pineda:2000sz}.

\par 
Writing the eigenstate of the pNRQCD hamiltonian at $O(m^0)$
in the ${\bf 3} \otimes {\bf 3}^{*}$ representation of SU(3) color,
which corresponds to the static quark-antiquark state,
as $| n \> \equiv | n; \vec{r}_{1},\vec{r}_{2} \>$,
the correlator of two color-electric field
strength operators
$E^{i}=F_{4i}$ ($i=1,2,3$), 
put on
$\vec{r}_{a}$ and $\vec{r}_{b}$ ($a,b=1,2$) in space
and separated $t=t_{1}-t_{2}$ in time, takes the form
\bea
 C(r,t) 
= \sum_{n=1}^{\infty} \< 0 | g E^{i}(\vec{r}_{a}) |n\>
\< n| g E^{j}(\vec{r}_{b})
| 0\> e^{-(\Delta E_{n0}(r))t}\;,
\label{eqn:correlator-infinite}
\eea
where $\Delta E_{n0}(r) \equiv E_{n}(r)-E_{0}(r)$ denotes
the energy gap with $E_{0}(r) =V^{(0)}(r)$.

\par
Then,
the $O(1/m)$ potential is given by
\bea
V^{(1)}(r) = - \frac{1}{2}\delta_{ij}\sum_{n = 1}^{\infty}
\frac{\< 0 | g E^{i} (\vec{r}_{1})| n\> 
\< n | g E^{j} (\vec{r}_{1})| 0\>}
{(\Delta E_{n0}(r) )^2} \; ,
\label{eqn:v1-spectralrep}
\eea
where two color-electric field strengths are
attached to one of the quarks.

\par
The spin-independent part of the $O(1/m^2)$ potential
is written as
\bea
V_{\rm SI}(r)
&=&
\frac{1}{m_{1}^2}
\left (\frac{1}{2} \{ {\vec{p}_{1}}^2, V_{p^2}^{(2,0)}(r) \}
+ \frac{1}{r^2} V_{l^{2}}^{(2,0)}(r)
\vec{l}_{1}^{\;2} +V_{r}^{(2,0)}(r) \right )  \nn\\
&&
+
\frac{1}{m_{2}^2}
\left ( \frac{1}{2} \{ {\vec{p}_{1}}^2, V_{p^2}^{(0,2)}(r) \}
+ \frac{1}{r^2} V_{l^{2}}^{(0,2)}(r)
\vec{l}_{2}^{\;2} +V_{r}^{(0,2)}(r) \right )  \nn\\
&&
+\frac{1}{m_{1}m_{2}}
\left (-\frac{1}{2} \{ \vec{p}_{1}\! \cdot \! \vec{p}_{2},
V_{p^{2}}^{(1,1)}(r)\}
-
\frac{1}{2r^2}
V_{l^2}^{(1,1)}(r)
(\vec{l}_{1}\cdot \vec{l}_{2}+
\vec{l}_{2}\cdot\vec{l}_{1})
+V_{r}^{(1,1)}(r)
\right )\;,
\eea
where $\vec{l}_{a}=\vec{r}\times \vec{p}_{a}$.
The functions specified 
by the subscripts $p^2$ and $l^2$ 
are related to the velocity-dependent potentials,
$V_{b}$, $V_{c}$, $V_{d}$ and $V_{e}$ defined
in Refs.~\cite{Barchielli:1986zs,Barchielli:1988zp}, by
\bea
&&
V_{{p}^{2}}^{(2,0)} (r)=
V_{{p}^{2}}^{(0,2)} (r)=
V_{d}(r)-\frac{2}{3} V_{e}(r) \; ,\quad
V_{{l}^{2}}^{(2,0)} (r)=
V_{{l}^{2}}^{(0,2)} (r)=
V_{e}(r) \;,\nn\\
&&
V_{{p}^{2}}^{(1,1)} (r)= 
-V_{b}(r) +\frac{2}{3}V_{c}(r) \;, \quad
V_{{l}^{2}}^{(1,1)}(r) = 
-V_{c}(r) \; ,
\eea
and then we have
\bea
&&
V_{b}(r)=  -\frac{2}{3} \delta_{ij}\sum_{n=1}^{\infty}
\frac{\<0| g E^{i} (\vec{r}_{1})| n\> 
\< n|g E^{j} (\vec{r}_{2}) |0 \>}{(\Delta E_{n0})^3}
\;,\\
&&
V_{c}(r)= 3 \left (\frac{r_{i} r_{j}}{r^2}-\frac{\delta_{ij}}{3}
\right )\sum_{n=1}^{\infty}
\frac{\<0| g E^{i} (\vec{r}_{1})| n\> 
\< n|g E^{j} (\vec{r}_{2}) |0 \>}{(\Delta E_{n0})^3}
\;,\\
&&
V_{d}(r)= \frac{1}{3} \delta_{ij}\sum_{n=1}^{\infty}
\frac{\<0| gE^{i}(\vec{r}_{1})| n\> 
\< n|g E^{j}(\vec{r}_{1}) |0 \>}{(\Delta E_{n0})^3}
\;,\\
&&
V_{e}(r)= -\frac{3}{2}
\left (\frac{r_{i} r_{j}}{r^2}-\frac{\delta_{ij}}{3}
\right )\sum_{n=1}^{\infty}
\frac{\<0| g E^{i} (\vec{r}_{1})| n\> 
\< n|g E^{j} (\vec{r}_{1}) |0 \>}
{(\Delta E_{n0})^3}
\;.
\eea
For $V_{d}$ and $V_{e}$ two field strengths are attached to
one of the quarks as for the $V^{(1)}$, while
for $V_{b}$ and $V_{c}$ two field strengths are attached to
the quark and the antiquark, respectively.

\par
Thus, once the matrix elements and the energy gaps
are known from the behavior of the field strength correlators
on the quark-antiquark source,
one can compute the potentials.

\section{Numerical procedures}

We work in Euclidean space in four dimensions on a
hypercubic lattice with lattice volume $V=L^3 T$ and 
lattice spacing $a$, where we impose
periodic boundary conditions in all directions.
We use the Polyakov loop correlation function (PLCF, 
a pair of Polyakov loops $P$ separated by a distance $r$)
as the quark-antiquark source
and evaluate the color-electric field strength 
correlators on the PLCF,
\bea
C(r,t) =\[ g E^{i} (\vec{r}_{a},t_{1}) g E^{j} 
(\vec{r}_{b},t_{2}) \]_{c}
=
\[ g E^{i} (\vec{r}_{a},t_{1}) g E^{j} (\vec{r}_{b},t_{2}) \]
\!- \! \[ g E^{i} (\vec{r}_{a})\]  \[ g E^{j} (\vec{r}_{b})\] \;,
\label{eqn:correlator}
\eea
using the multi-level algorithm~\cite{Koma:2006si,Koma:2006fw},
where the double bracket represents the ratio of expectation values
$\[\cdots \] =  \< \cdots \>_{PP^{*}} /  \< P P^{*}(r)\>$, while
$\< \cdots \>_{PP^{*}}$ means that 
the color-electric field is connected to 
the Polyakov loop in a gauge invariant way.
The subtracted term on the r.h.s. of Eq.~\eqref{eqn:correlator}
can be nonzero as the 
color-electric field is even under $\mathit{CP}$ transformations.
The spectral representation of Eq.~\eqref{eqn:correlator} 
derived with transfer matrix theory reads~\cite{Koma:2006fw}
\bea
C(r,t) 
=
2 \!\! \sum_{n = 1}^{\infty} 
\< 0  | g E^{i}(\vec{r}_{a}) | n \> 
\< n | g E^{j} (\vec{r}_{b}) | 0 \>
e^{- (\Delta E_{n0})T/2} 
\cosh ((\Delta E_{n0})(\frac{T}{2} -t)) \! 
+ \!O(e^{-(\Delta E_{10})T})\;  ,
\label{eqn:correlator-spectralrep}
\eea
where the last term represents terms of 
exponential factors equal to or smaller than
$\exp (-(\Delta E_{10})T)$, which
can be neglected for a reasonably large $T$.
Thus, once Eq.~\eqref{eqn:correlator} is evaluated
via Monte Carlo simulations, we can
fit the matrix element
$\< 0 | g E^{i}(\vec{r}_{a}) |n \>\< n | g E^{j}(\vec{r}_{b}) |0 \>$
and the energy gap $\Delta E_{n0}$ 
in Eq.~\eqref{eqn:correlator-spectralrep},
both of which are finally
inserted into Eq.~\eqref{eqn:v1-spectralrep}, etc.
It is clear that 
Eq.~\eqref{eqn:correlator-spectralrep} is reduced to
the form like Eq.~\eqref{eqn:correlator-infinite}
in  the infinite volume limit $T\to \infty$.

\par
We define 
the lattice color-electric field operator,
$g a^2 E^{i}(s)$, from the traceless part of
$[ U_{4i}(s)-U^{\dagger}_{4i}(s)]/(2i)$
with two-leaf modification (an average of $F_{4i}(s)$
and $F_{4i}(s-\hat{i})$), where $U_{\mu\nu}(s)$ is a
plaquette variable defined on the site $s$.
We multiply the Huntley-Michael
factor~\cite{Huntley:1986de} on the PLCF,
$Z_{E}$~\cite{Koma:2006fw},
to the lattice color-electric field 
to cancel the self energies at least at $O(g^2)$.

\par
We point out several
advantages of our procedure which enable us to reduce
numerical errors.
In earlier studies of the relativistic corrections on the lattice,
the Wilson loop has been used as the quark-antiquark source,
since the corrections have been expressed
as the integral of the field strength correlators on 
the Wilson loop with respect to the relative temporal distance
between two field strength operators, $t$ 
(see, e.g.~\cite{Eichten:1979pu}).
They are, in principle, measurable on the lattice,
and the result is reduced to
Eq.~\eqref{eqn:v1-spectralrep} once the spectral 
decomposition is applied by using transfer matrix theory,
and the temporal size of the Wilson loop is taken to infinity.
In practice, however, 
the integration of the field strength correlator 
and the extrapolation of the temporal size of 
the Wilson loop to infinity cause systematic errors.
In contrast, we can avoid these systematic errors
once the field strength correlator is determined accurately,
since we directly evaluate the matrix elements and the energy gaps.
The use of the PLCF as the quark-antiquark source 
and its spectral representation allows us to take into account
the finite-$T$ effect automatically in the fit.
We should note however that if one uses the commonly employed 
simulation algorithms, it is almost impossible to evaluate the 
field strength correlators on the PLCF, or the PLCF itself, at 
intermediate distances with reasonable computational effort,
since the expectation value of the PLCF at zero temperature
is smaller by several orders of magnitude than that of the
Wilson loop.
However this problem is solved by employing the
multi-level algorithm~\cite{Luscher:2002qv}.
For details of our implementation, 
see~\cite{Koma:2006fw}.

\section{Numerical results}

\begin{figure}[hbt]
\includegraphics[width=7cm]{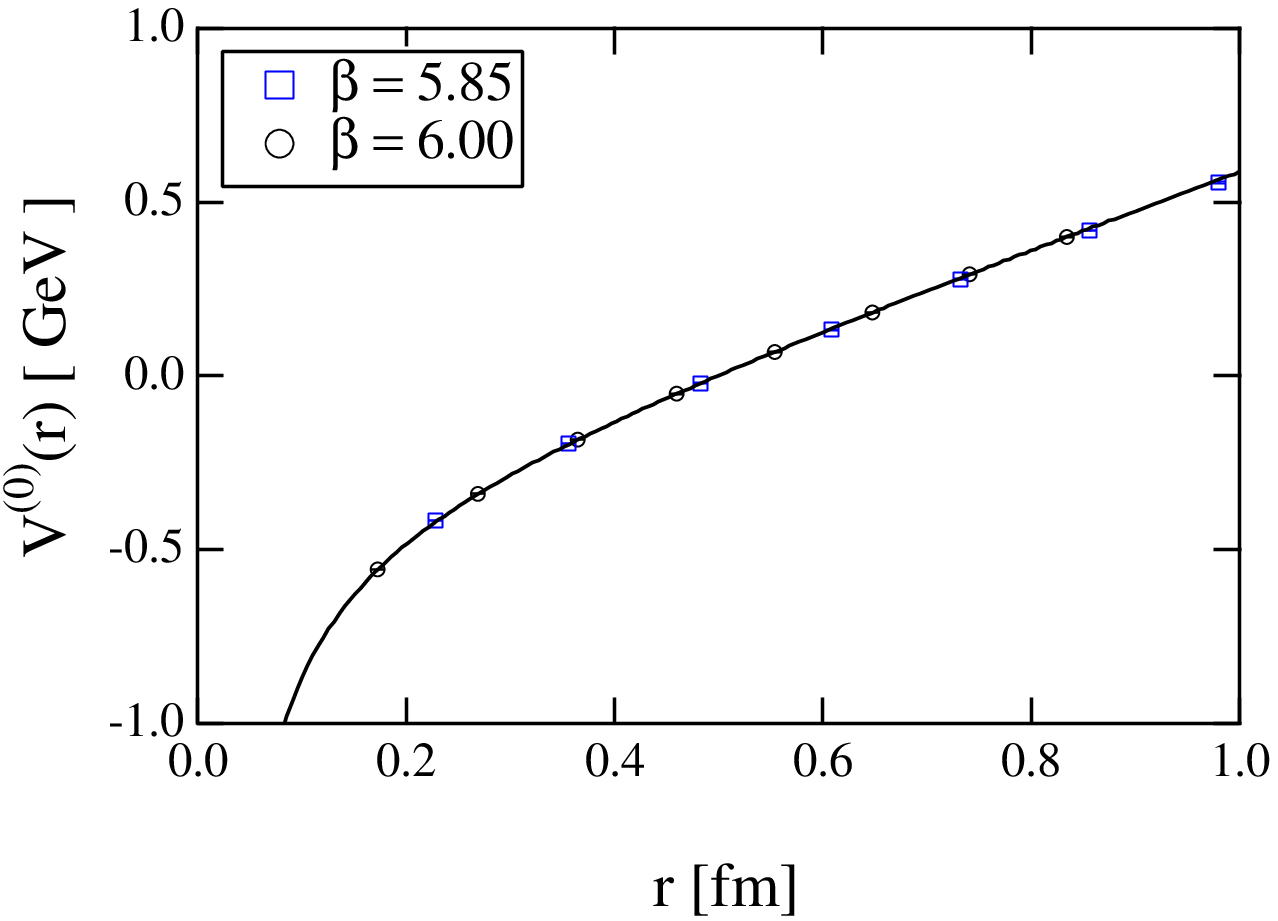}  
\includegraphics[width=7cm]{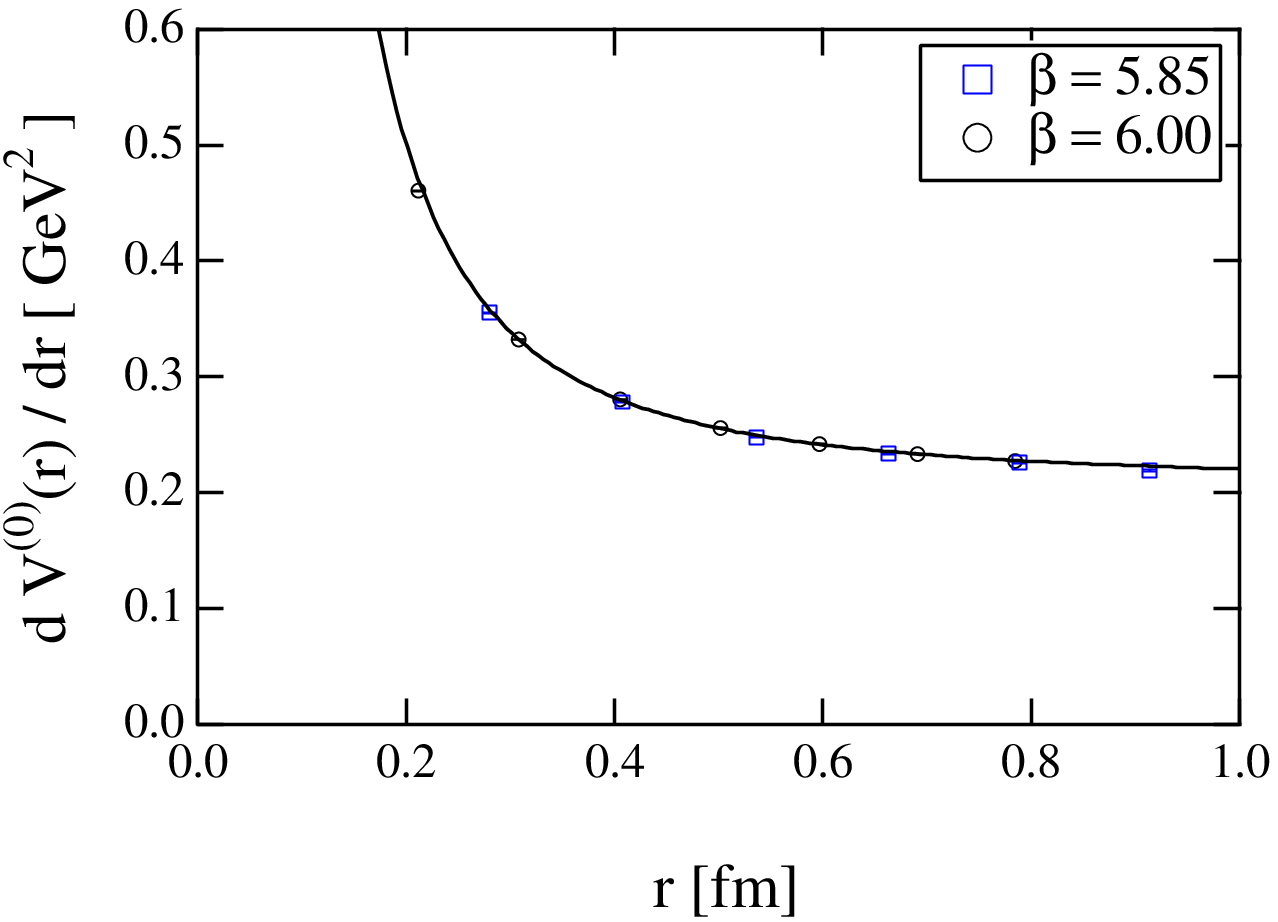}
\caption{Static potential $V^{(0)}(r)$ and the 
force $dV^{(0)}(r)/dr$
as a function of $r$, which is improved with
treelevel perturbation theory. 
The potentials of different $\beta$ values
are normalized at $r=0.5$~fm.}
\label{fig:pot-0}
\end{figure}

We carry out simulations using
the standard Wilson gauge action in SU(3) lattice gauge theory.
We summarize our simulation parameters in Table~\ref{tbl:para}.
The lattice spacing $a$ is set from the Sommer scale
$r_{0}=0.5$~fm.
For a reference we compute the static potential and the force
from the PLCF,
\bea
V^{(0)}(r) = -\frac{1}{T} \ln 
\< PP^*(r)\>  + O(e^{-(\Delta E_{10})T})\;, \quad
\frac{dV^{(0)}(r)}{dr} = \frac{V^{(0)}(r)-V^{(0)}(r-a)}{a}\;, 
\eea
which are shown in Fig.~\ref{fig:pot-0}.
The fitting of the potential data at $\beta=6.00$ 
to the functional form in
Eq.~\eqref{eqn:pot-0} yields
$c= 0.2808(5)$, $ \sigma a^2=0.0468(1)$ and $\mu a =0.7301(4)$ with
$\chi^2/N_{\rm df}=3.5$.
We note that the large value of $\chi^{2}/N_{\rm df}$ just reflects the fact that
the Coulombic coefficient is not strictly constant as a function of $r$,
which will be clear once the second derivative of the potential is
investigated~\cite{Luscher:2002qv,Koma:2006fw},
but we ignore this effect for the moment.

\begin{table}[hbt]
\caption{Simulation parameters used in this study.
$N_{\mathrm{tsl}}$ is the number of time slices in a sublattice
and $N_{\mathrm{iupd}}$ the number of internal update within a
sublattice, both are parameters for the multi-level 
algorithm.}\vspace*{0.2cm}
\centering
\begin{tabular}{cccccc}
    \hline
    $\beta=6/g^2$& $a$ [fm] & $(L/a)^3 (T/a)$
    & $N_{\mathrm{tsl}}$ & $N_{\mathrm{iupd}}$ & 
    $N_{\mathrm{conf}}$ \\
    \hline
    5.85 & 0.123 &  $18^{3}24$ &  3 & 50000 & 100\\
    6.00 & 0.093 &  $24^{3}32$ &  4 & 50000 & 45 \\
    \hline 
\end{tabular}
\label{tbl:para}
\end{table}

\par
\begin{figure}[!t]
\includegraphics[width=7cm]{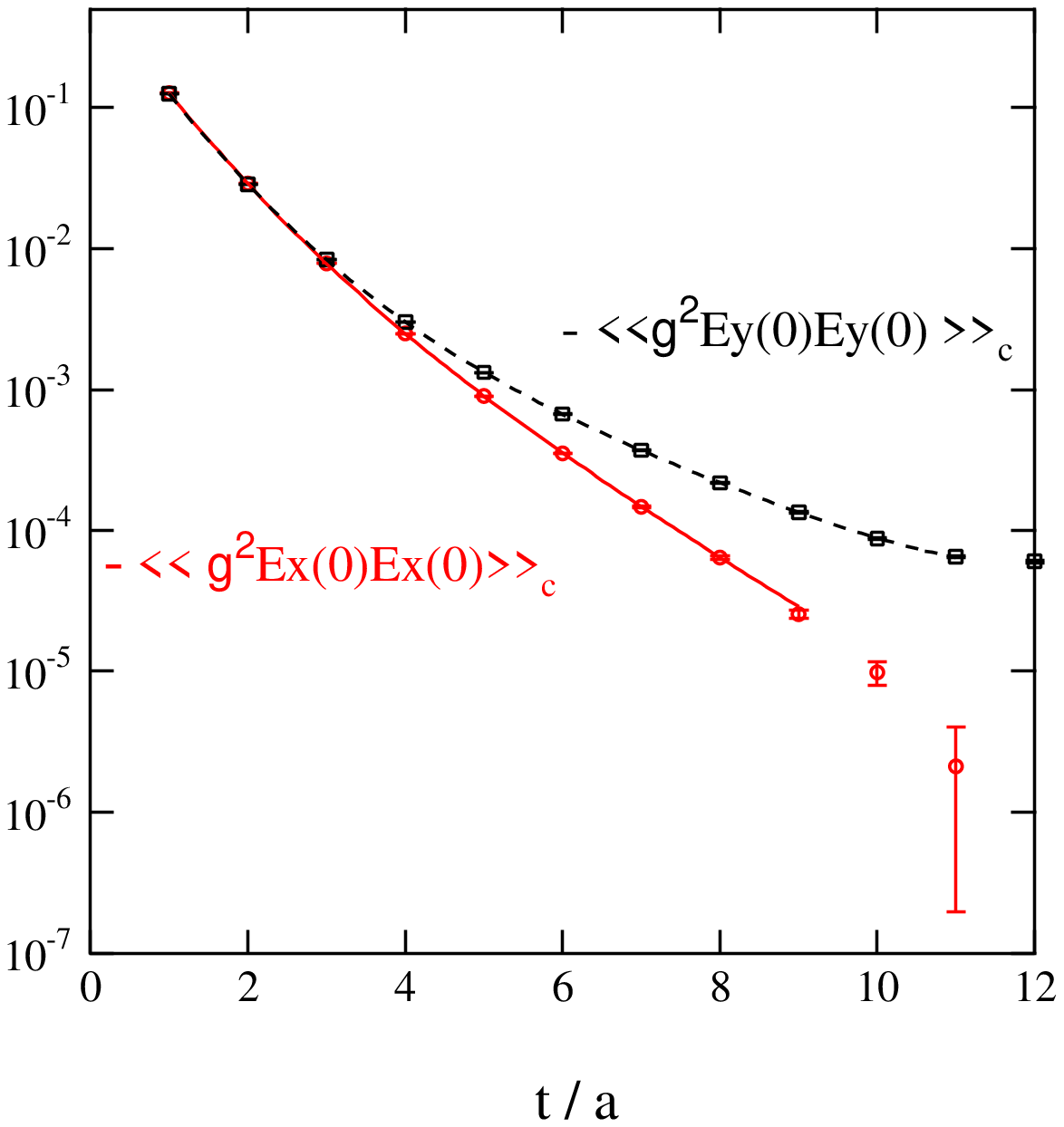}
\includegraphics[width=7cm]{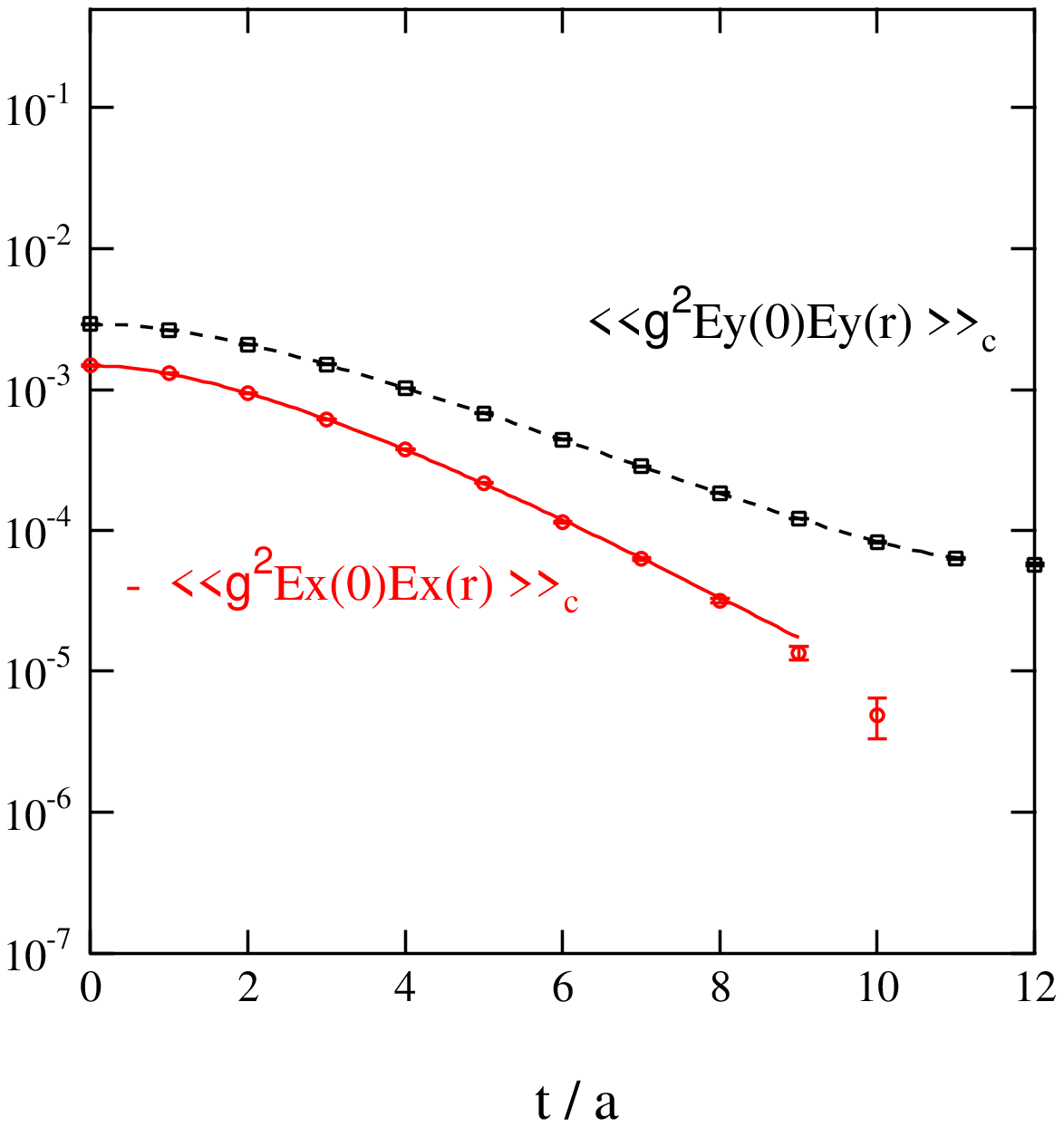}
\caption{Color-electric field strength correlators at $\beta=5.85$
on the $18^3 24$ lattice for $r/a=5$.}
\label{fig:efsc}
\end{figure}

In Fig.~\ref{fig:efsc}, we show
the typical behavior of 
the longitudinal and the 
transverse components of the color-electric
field strength correlators, where
the quark-antiquark axis has been taken along the $x$ axis.  
We select the data
at $r/a=5$, $\beta=5.85$
as an example.
Statistical errors of the correlators are small enough to
determine the matrix elements and the energy gaps
with the fit based on Eq.~\eqref{eqn:correlator-spectralrep}.
It is evident that Eq.~\eqref{eqn:correlator-spectralrep}
describes the data very well.
Here, since it is impossible to determine the matrix elements
and the energy gaps for all $n\geq 1$ with the limited 
data points, we truncated expansion in 
Eq.~\eqref{eqn:correlator-spectralrep} at a certain $n =n_{\rm max}$.
The validity of the truncation was monitored
by looking at the reduced $\chi^2$ defined with the 
full covariance matrix.
In all cases we found that $n_{\rm max}=3$ 
gave the best result.

\begin{figure}[!t]
\includegraphics[width=7cm]{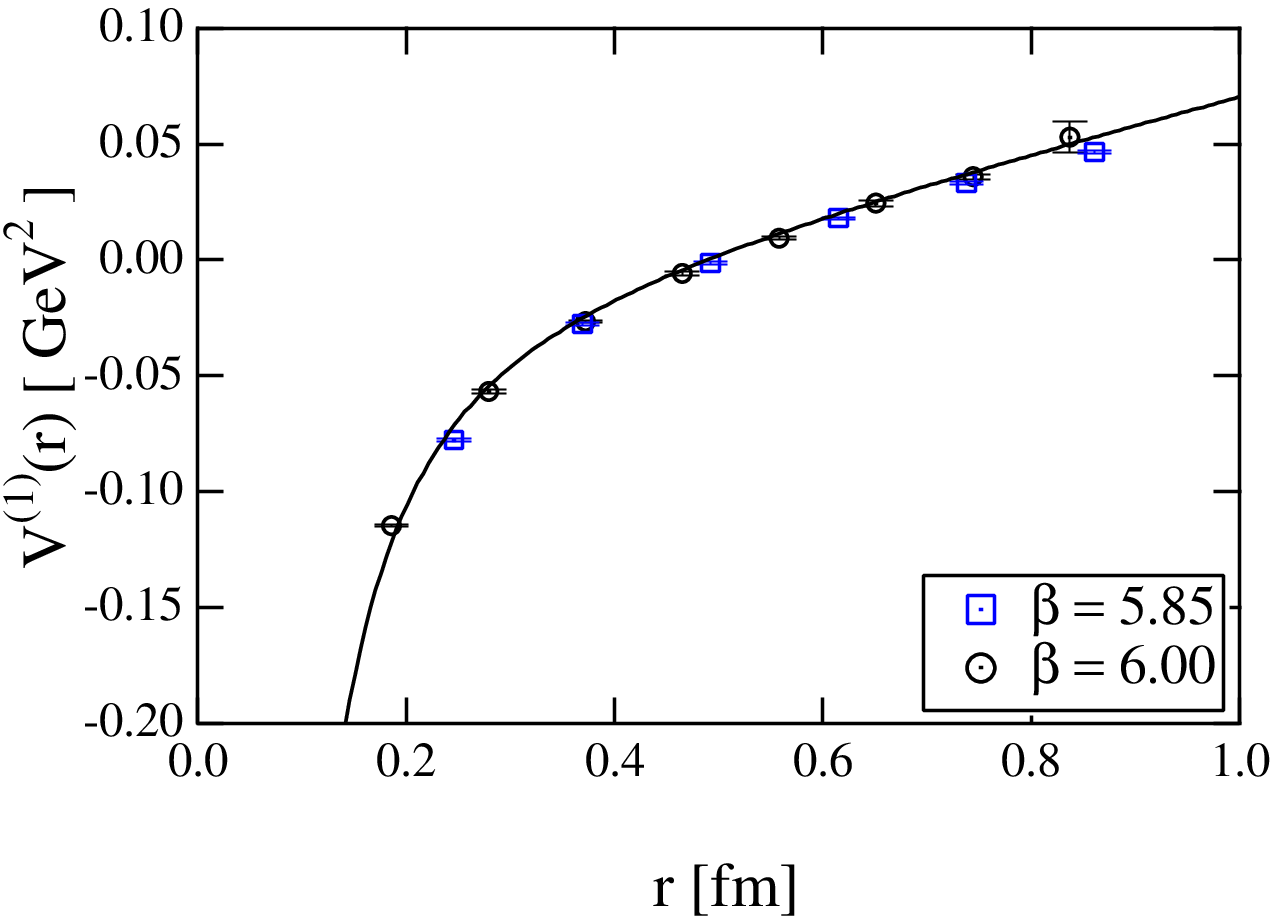}  
\includegraphics[width=7cm]{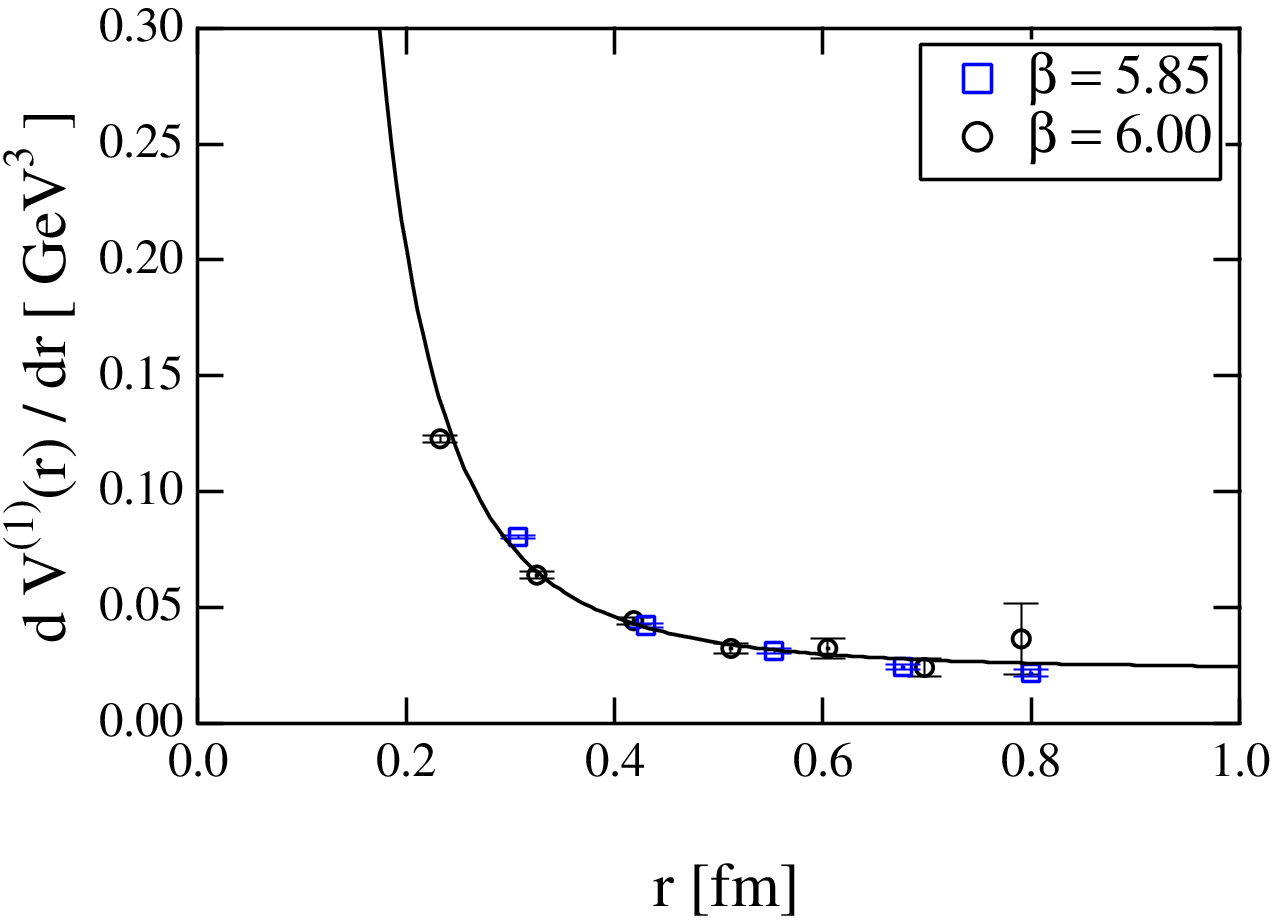}
\caption{The $O(1/m)$ potential and the  
force. }
\label{fig:pot-1m}
\end{figure}

\par
\begin{figure}[!t]
\includegraphics[width=7cm]{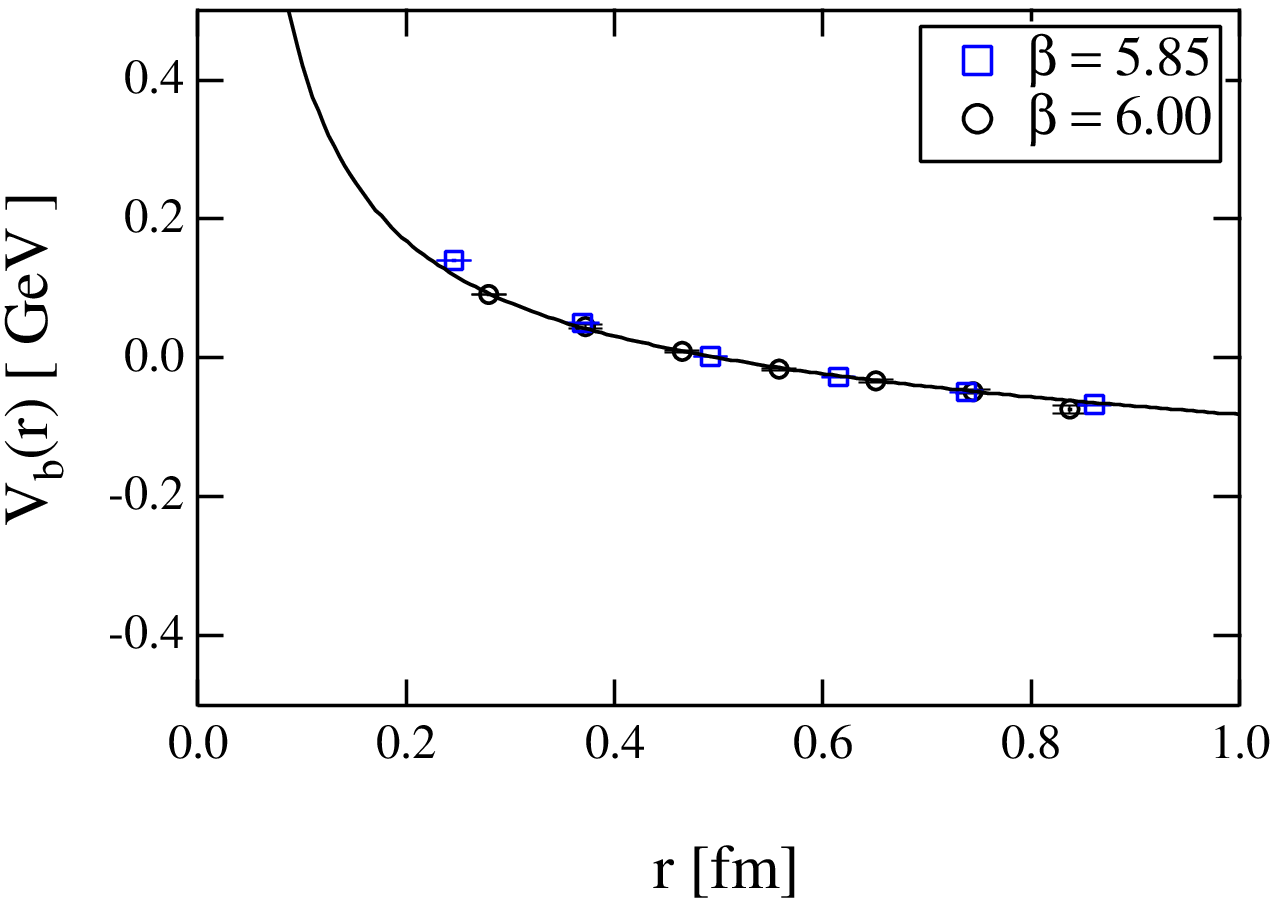}
\includegraphics[width=7cm]{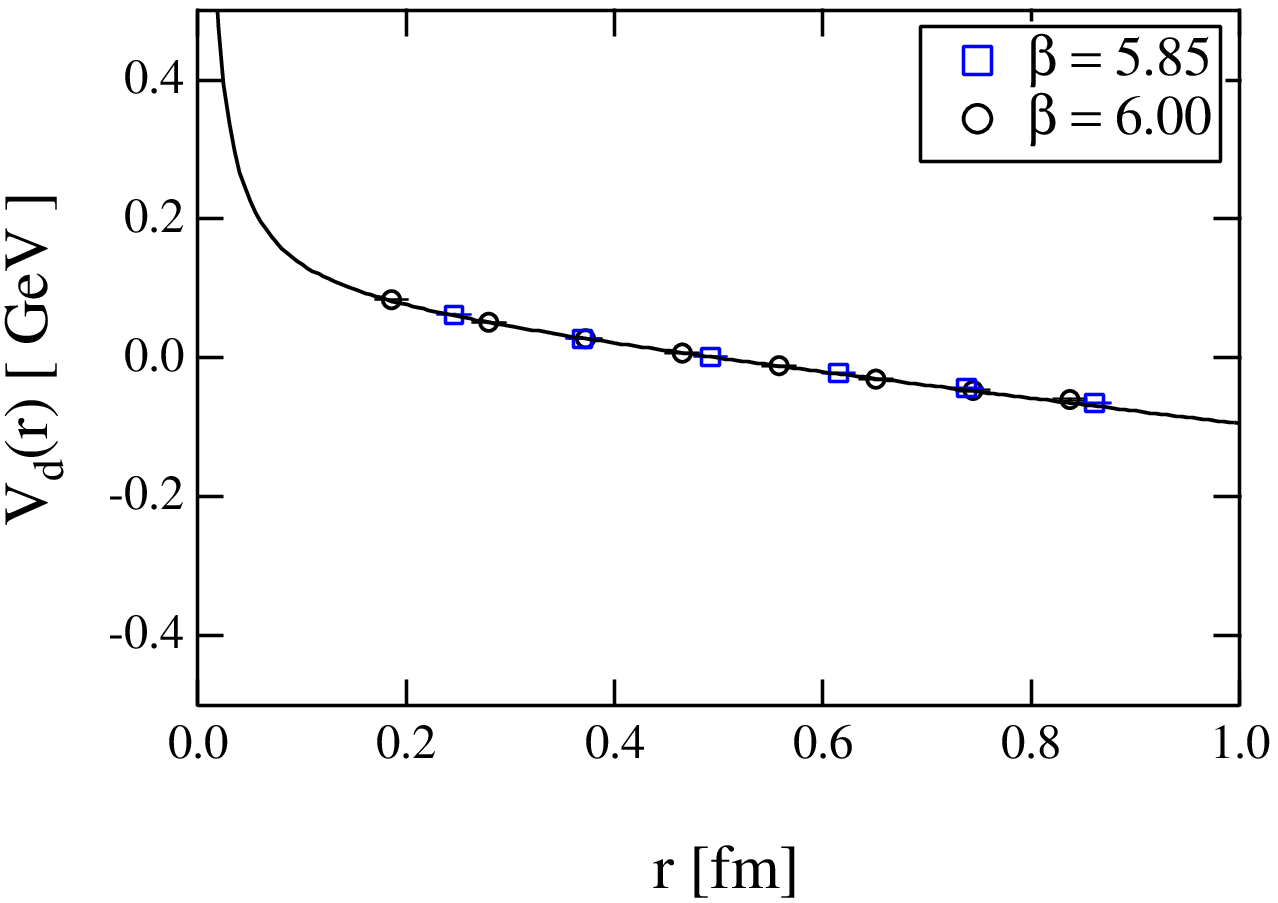}\\
\includegraphics[width=7cm]{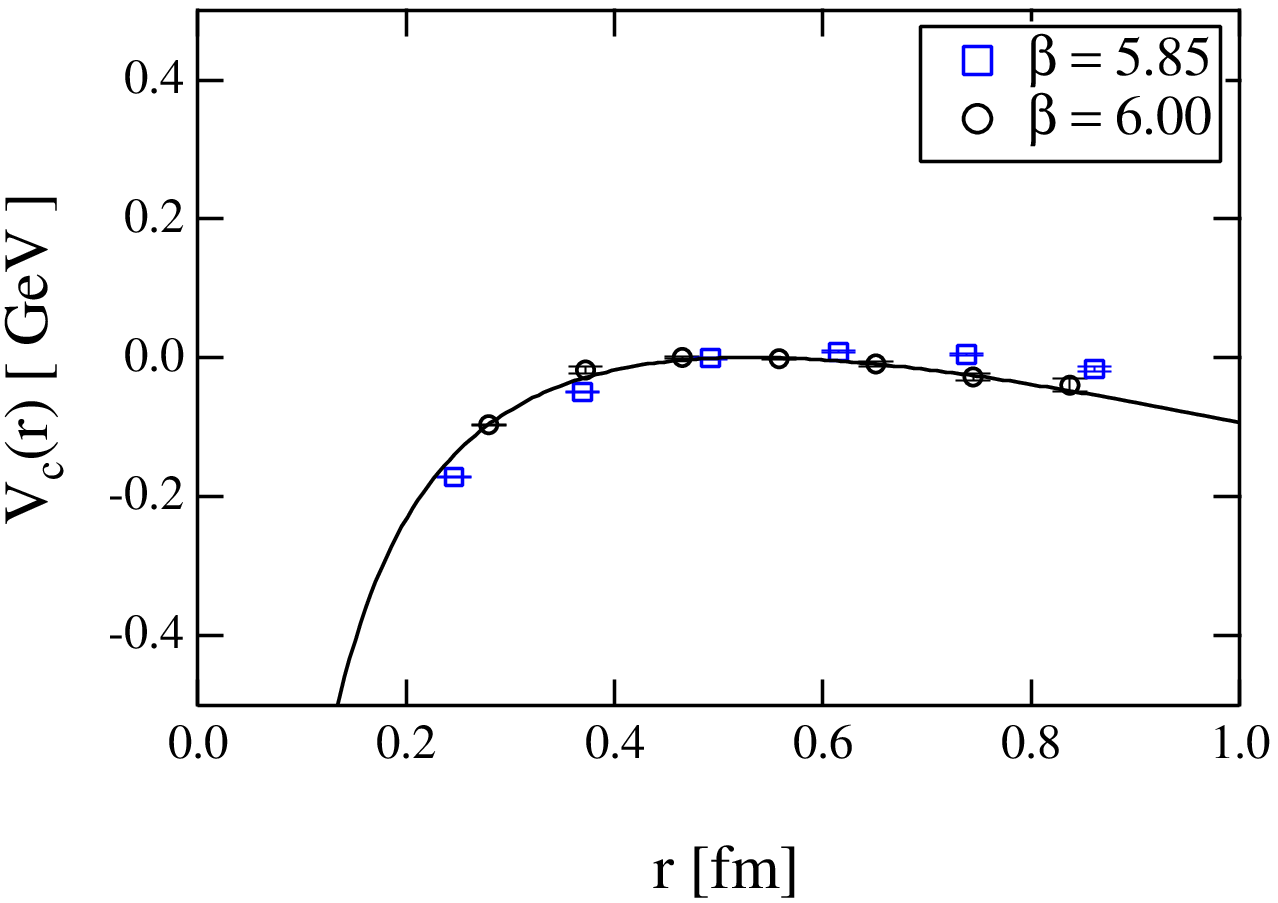}
\includegraphics[width=7cm]{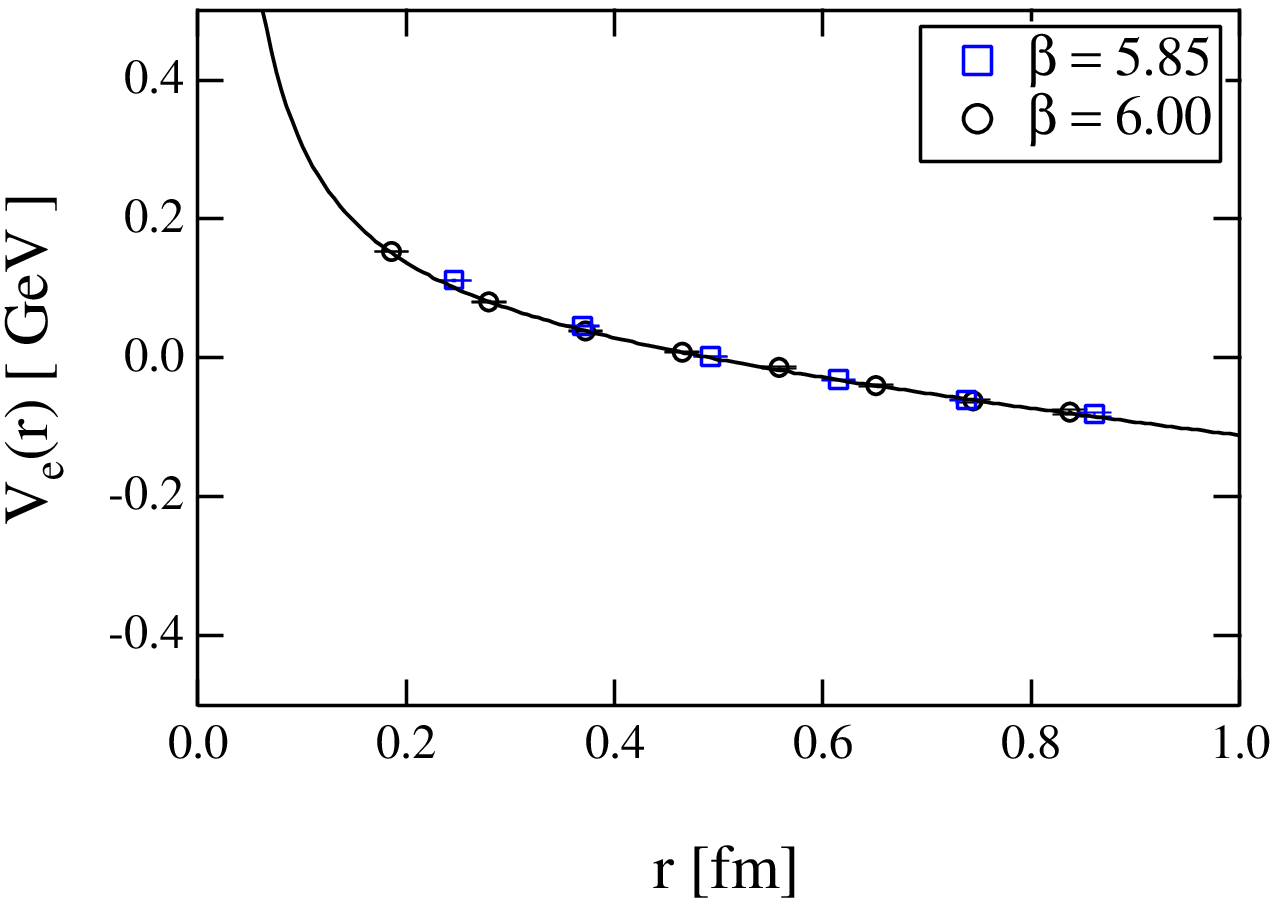}
\caption{The  $O(1/m^2)$ velocity-dependent potentials.}
\label{fig:pot-vel}
\end{figure}

We then computed
the $O(1/m)$ potential and the $O(1/m^2)$ 
velocity-dependent
potentials, which are plotted in Figs.~\ref{fig:pot-1m}
and~\ref{fig:pot-vel}, respectively.
Note that each potential 
contains a constant contribution, which depends on $\beta$.
Here we normalized the potentials at 
$r=0.5$~fm by assuming 
perfect scaling behavior.
This assumption is justified at intermediate distance, where
the data at different lattice spacings fall into a smooth curve.
Small discrepancies at short distances, however,
cannot be avoided in the present simulation.
The data at short distances are sensitive to the way 
of discretization and the definition of the renormalized
color-electric field operator.

\par
The functional form of the $O(1/m)$ potential 
is not yet established nonperturbatively.
In our previous study~\cite{Koma:2006si},
we found that a $1/r$ function
describes the data well, but this result was
valid only up to $r=0.6$~fm.
Now we have further long distance data up to 
distances of $r=0.9$~fm.
We empirically examined various functional
forms and found
that,  if we include the data $r>0.6$~fm,
the $1/r$ function is not 
supported by the fit, while
the perturbative $1/r^2$ function with 
the linear term can fit the data well.
Using the functional form
\bea
V_{\rm fit}^{(1)}(r) = -\frac{c'}{r^{2}} +\sigma' r +\mu' \;,
\eea
we obtain $c'=0.090(5)$, $\sigma' a^{3}=0.0024(1)$, 
and $\mu' a^{2}=0.389(1)$
with $\chi^{2}/N_{\rm df}=0.68$ for the data at $\beta=6.00$,
where the distance $r/a=2$ is omitted.
In Fig.~\ref{fig:pot-1m}, we also plot the force of the $O(1/m)$
potential defined by
$\frac{d V^{(1)}(r)}{dr}=(V^{(1)}(r)-V^{(1)}(r-a))/a$.
We find that the fit of the force to the function 
$\frac{d V_{\rm fit}^{(1)}(r)}{dr}=\frac{2c'}{r^3}+\sigma'$
gives $c'=0.095(5), \sigma'=0.0024(1)$ with $\chi^{2}/N_{\rm 
df}=0.62$,
which is consistent with the fit result of the potential.

By taking into account the masses $(1/m_{1} +1/m_{2})$
in Eq.~\eqref{eqn:pnrqcd-hamiltonian}, we may estimate 
the correction to the string tension in the static
potential.
For charmonium, $m_{c}=1.3$~GeV, we find 
$(2/m_{c})\sigma' = 0.179 $~GeV/fm, which is compared to
$\sigma=1.07(1)$~GeV/fm. 
The correction is about~17~\%.
For bottomonium, $m_{b}=4.7$~GeV, we find
$(2/m_{b})\sigma' = 0.049 $~GeV/fm, so that the correction is
about~5~\%.

\par
Next, we may characterize the functional form 
of the velocity-dependent potentials.
Here, motivated by the minimal area law (MAL) 
model~\cite{Barchielli:1986zs,Barchielli:1988zp},
we fit the potentials to the same functional form 
as the static potential.
Again, we omit the point at  $r/a=2$ from the analysis
of the data taken at $\beta=6.00$.
The fitting results are summarized in Table~\ref{tbl:fit-velocity}.
The global structure of the data seems to
be well described by the fitting
function as seen in Fig.~\ref{fig:pot-vel},
though $\chi^{2}/N_{\rm df}$ may be relatively large.

\begin{table}[!t]
    \centering
    \caption{Fit result of the velocity dependent potential
    to the function $V_{\rm fit} (r)= -c/r + \sigma r +\mu$.}
    \vspace*{0.2cm}
\begin{tabular}{|c|l|l|l|c|}
\hline
    & \multicolumn{1}{c|}{$c$}
    & \multicolumn{1}{c|}{$\sigma a^2$ }
    &  \multicolumn{1}{c|}{$\mu a$}   & $\chi^{2}/N_{\rm df}$ \\
    \hline
$V_{b}$ & $-0.25(2)$ & $-0.003(1)$& $-0.08(1)$ & 1.5\\
$V_{c}$ & $ \phantom{-}0.61(4)$ & $-0.019(2)$ & $\phantom{-}0.08(2)$ & 3.0 \\
$V_{d}$ & $-0.042(4)$ &$ -0.0076(3)$ & $-0.187(2)$ & 4.3\\
$V_{e}$ & $ -0.156(4)$& $-0.0069(3)$ & $-0.017(2)$ & 4.6\\
\hline
\end{tabular}
\label{tbl:fit-velocity}
\end{table}

\par
\begin{figure}[!t]
\includegraphics[width=7cm]{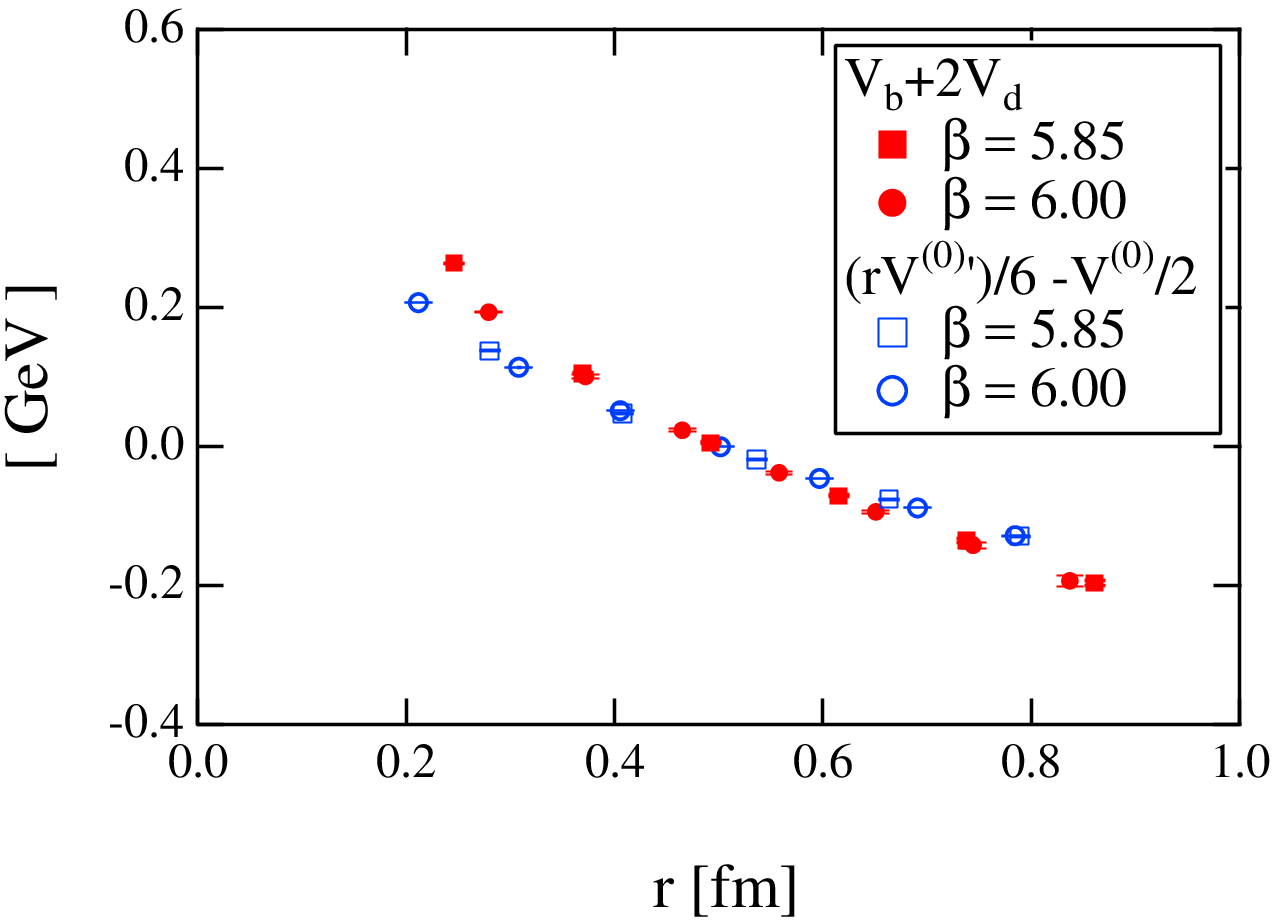}
\includegraphics[width=7cm]{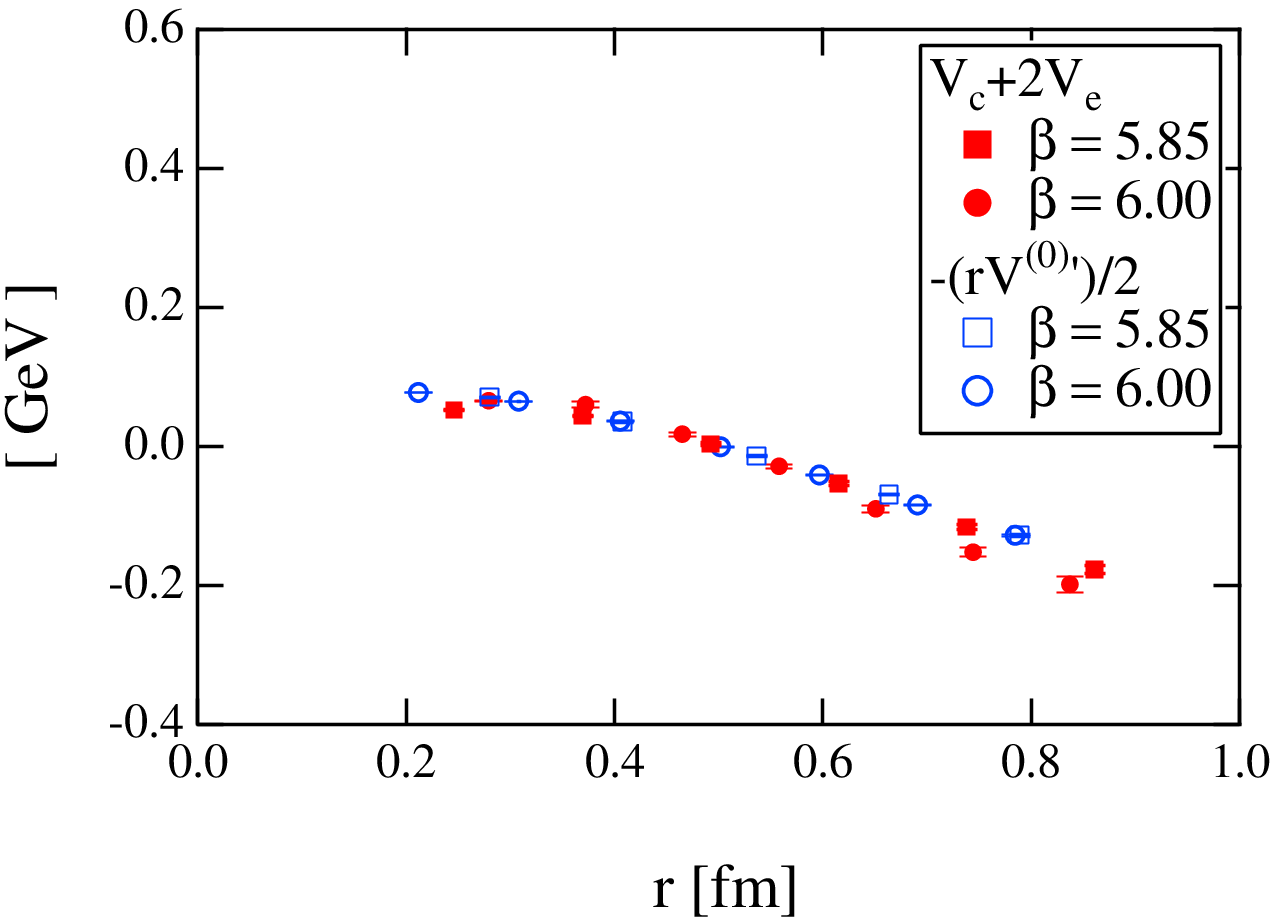}
\caption{Test of the BBMP relations in Eq.~(4.3).}
\label{fig:pot-vel-bbmp}
\end{figure}

Finally we examine some nonperturbative relations which
connect the velocity-dependent potentials
to the static potential~\cite{Barchielli:1986zs,Barchielli:1988zp},
which are often called the BBMP relations,
\bea
V_{b}(r)+2V_{d}(r) = -\frac{1}{2}V^{(0)}(r)
+\frac{r}{6}\frac{dV^{(0)}(r)}{dr}
\;,\quad 
V_{c}(r)+2V_{e}(r) = -\frac{r}{2}\frac{dV^{(0)}(r)}{dr} \;.
\label{eqn:bbmp}
\eea
These relations are 
derived by exploiting the exact Poincar\'e  invariance of the
field strength correlator, 
and hence one expects corrections due to lattice artifacts, 
which would vanish in the continuum limit.
These relations can be regarded as an extension of the Gromes 
relation~\cite{Gromes:1984ma} 
for the $O(1/m^{2})$ spin-dependent potentials. 
In Fig.~\ref{fig:pot-vel-bbmp}, we show the result,
where the constant contribution of the potentials
is normalized at $r=0.5$~fm.
The BBMP relations seem to be satisfied, though we see a
small discrepancy especially at short distances.

\section{Summary}

We have investigated the relativistic corrections to the
static potential, the $O(1/m)$ potential and the  $O(1/m^{2})$
velocity-dependent potentials, in SU(3) lattice gauge theory.
They are important  ingredients of the pNRQCD hamiltonian
for heavy quarkonium.

\par
By evaluating the color-electric field strength correlator
on the PLCF with the multi-level algorithm, 
and exploiting the spectral representation of the correlator,
we have obtained a very clean signal for
these potentials up to $r=0.9$~fm.
The  $O(1/m)$ potential contains a linearly rising 
nonperturbative contribution.
The $O(1/m^{2})$ velocity-dependent potentials 
are non-vanishing at long distances.

\par
All potentials at different $\beta$ values, 
normalized at  $r=0.5$~fm, show a reasonable scaling behavior.
The BBMP relations are apparently satisfied 
around $r\simeq$ 0.5 fm.
Although we have applied the Huntley-Michael prescription
in the present study to remove the self-energy contributions
of the field strength operator,
a more systematic, non-perturbative renormalization procedure of 
field strength operators is highly desirable.
Since the statistical errors of the potentials are
reduced significantly
owing to the multi-level algorithm,
we now face such a delicate problem.

\par
The comparison with various 
models~\cite{Barchielli:1988zp,Brambilla:1996aq}
and 
phenomenology~\cite{Ebert:1997nk,Ebert:1999xv,Ebert:2002pp} 
are of course to be done. 
In particular, it is quite interesting to
examine the effect of the $O(1/m)$ potential 
on the spectrum as this is 
the leading-order relativistic correction in
the $1/m$ expansion.

\section*{Acknowledgments}

The main calculation has been performed on the NEC-SX8 
at Research Center for Nuclear Physics (RCNP), 
Osaka University, Japan.

\end{document}